\documentclass[preprint,preprintnumbers,amsmath,amssymb,amsfonts]{revtex4}
\usepackage[a4paper,left=30mm,right=10mm,top=20mm,bottom=20mm,nohead]{geometry}
\usepackage{bm}
\begin{document}
\title{Geometrical phase for a three-dimensional anisotropic quantum well}
\author{V. A. Geyler}
\email{geyler@mrsu.ru}
\author{A. V. Shorokhov}
\email{alex.shorokhov@mail.ru} \affiliation {Mordovian State
University, 430000 Saransk, Russia}
\begin{abstract}
A three-dimensional anisotropic quantum well placed in an
adiabatically precessing uniform magnetic field is considered and
an explicit formula for the Berry phase is obtained. To get the
Berry phase, a purely algebraic algorithm of reducing a quadratic
Hamiltonian to the canonical form via symplectic transformations
of the phase space is presented.
\end{abstract}
 \maketitle
\newpage
\section{Introduction}
In the classical work [\onlinecite{Ber84}], M.Berry has shown that
if an evolution of a physical system is cyclic and adiabatic the
geometric phase (so-called Berry phase) is accumulated on the wave
function in addition to the usual dynamical phase. The Berry phase
has a purely geometric nature and reflects geometric properties of
the parameter space of the system. B.Simon [\onlinecite{Sim83}]
has given an interpretation  of the Berry phase as due to holonomy
in a line bundle over the parameter space.

The space of values of the magnetic induction as a parameter space
for the considered system is frequently dealt with.  As a rule,
for a magnetic field, the Berry phase is studied in the two cases:
(1) a potential well is transported along a closed loop in the
usual configuration space [\onlinecite{Ber97,Zhu00,Exn00}]; (2)
spin particles interact with a time-depended magnetic field
[\onlinecite{Bit87,Ric88,Wei90}]. The most part of theoretical
studies is devoted to the evolution of the spin in a magnetic
field; in this case explicit formulae are easy to obtain using
finite-dimensional algebra. However, the Berry phase can arise in
the case of a spin-less particle interacting with a precessing
magnetic field [\onlinecite{Fer92}].

The aim of this paper is to obtain an explicit formula for the
Berry phase in the case of a three-dimensional  anisotropic
quantum well placed in a precessing magnetic field. The classical
Hamiltonian of such a system is a quadratic form of momenta and
positions. The considered problem has attracted attention of
physicists since the  1980's. In particular,  M.Berry
[\onlinecite{Ber85}] has calculated the geometric phase for a
one-dimensional quadratic potential with variable coefficients. As
for now, the most general result for quadratic Hamiltonians is
contained in [\onlinecite{Fer92}] where the following simple case
has been studied: a quadratic potential has an axial symmetry with
respect to the $z$-axis and a magnetic field precesses around
$z$-axis. These assumptions reduce the dimension of the problem
and significantly simplify calculations. We investigate the most
general problem in the case in which the frequencies of the
parabolic potential are mutually different and the magnetic field
is arbitrarily directed.

The significant peculiarity of a quadratic quantum Hamiltonian
lies in the fact that the Feynman integrals for their propagators
can be calculated explicitly. What is more, the problem of
calculating these integrals is reduced to finding the canonical
form  for a classical counterpart of the Hamiltonian via
symplectic transformations of the phase space; in our case this
form is a sum of squares of momenta and positions). Finding the
canonical form is a standard (but as a rule very cumbersome)
problem of linear algebra. For convenience of readers,  in the
next section we present an algorithm of finding the classical
(symplectic) and quantum (unitary) transformations that reduce the
corresponding Hamiltonian to the canonical form.

It is important to stress that we obtain our results using only
simple methods of linear algebra. It is strange enough that in the
similar situations, more complicated methods are used in the
physical literature; in particular, the method of the Bogolubov
canonical transformations [\onlinecite{Fer87}] or other indirect
methods. Note another advantage of the method we use. To find
various physical quantities one needs the matrix elements of the
corresponding operators. However, a direct calculation of matrix
elements can be a complicated problem. Our method allows to
resolve this problem using only simple calculations from linear
algebra. As an example we can mention the papers
[\onlinecite{Gey01}] and [\onlinecite{Mar02}] where the hybrid and
hybrid-phonon resonances for an asymmetric quantum well in a
magnetic field have been studied by means of this approach.

\section{Preliminaries}
The advantage of a description of mechanical systems with the help
of the Hamilton formalism is conditioned by the following main
reason: a choice of the generalized coordinates is restricted by
no conditions (any one of quantities defining the state of a
system may be the generalized coordinates.) What is more it is
possible to choice such a class of transformations of phase
coordinates that includes the transformation of the all initial
phase coordinates $p_i,q_i$ to the new ones $P_i,Q_i$ (here
$i=1,2...n$; $n$ is the dimension of the configuration space)
because momenta and positions play the same role in Hamilton
equations
\begin{equation}
\label{1}
Q=Q(p,q);P=P(p,q).
\end{equation}
However there is a restriction. The transformation (\ref{1}) is
permissible (canonical) if and only if it conserves the Poisson
brackets. The equivalent criterion for the transformation
(\ref{1}) to be canonical is the equality to the exact
differential the 1-form $pdq-PdQ$. This 1-form is the exact
differential of the generating function of the canonical
transformation (\ref{1}).
\begin{equation}
\label{2} dS=pdq-PdQ.
\end{equation}
Note that if we know the generating function we can always find
the functions determing the transformation (\ref{1}).

Through the paper we use the notation
\begin{equation}
\label{4} [ x_1|x_2]=-\langle p_1|q_2\rangle+\langle
p_2|q_1\rangle
\end{equation}
for the standard symplectic product in the phase space
$\mathbb{R}^{2n}=\mathbb{R}^n_q\times\mathbb{R}^n_p $\,; here
$x_j=(p_j,q_j)$, $(j=1,2)$.

Recall that a linear transformation
\begin{equation}
\label{5}
\left(\begin{array}{c}
p \\ q
\end{array}
\right)=L
\left(\begin{array}{c}
P \\ Q
\end{array}
\right).
\end{equation}
is canonical (i.e., preserves the Poisson brackets) if and only if
the matrix $L$ is symplectic:
\begin{equation}
\label{6} [ Lx_1|Lx_2]\equiv[ x_1|x_2].
\end{equation}
It is convenient to write the matrix $L$ in the form
\begin{equation}
\label{7}
\left(\begin{array}{cc}
L_1 & L_2 \\
L_3 & L_4
\end{array}\right)\,.
\end{equation}
The following commutation relations between components of the
matrix $L$ follow immediately from the fact that the matrix $L$ is
symplectic
\begin{equation}
\label{8} L_1^*L_4-L_2^*L_3=E\,, \ \ L_3^*L_4-L_4^*L_3=0\,, \ \
L_1^*L_2-L_2^*L_1=0\,,
\end{equation}
where $E$ is the unit matrix.

Denote by $S$ the generating function for the transformation $L$
(this means by definition that $dS=pdq-PdQ$); we will suppose that
$S$ is a function of $p$ and $P$. Then
\begin{equation}
\label{3} Q=-\frac{\partial S(p,P)}{\partial P}\,,\ \
q=\frac{\partial S(p,P)}{\partial p}\,.
\end{equation}
In this case the generating function of the canonical
transformation (\ref{5}) is expressed in terms of $L_i$ by the
following formula (we assume that $|L_2|\neq0$)
\begin{equation}
\label{9} S(p,P)=-\langle P|L_2^{-1}p\rangle+\frac{1}{2}\langle
P|L_2^{-1}L_1P\rangle+\frac{1}{2}\langle p|L_4L_2^{-1}p\rangle.
\end{equation}
According to the Dirac-Fock theorem [\onlinecite{Dir64}]  the
classical transformation (\ref{5}) corresponds to a quantum
transformation $U:\,L^2(\mathbb{R}^n_P)\rightarrow
L^2(\mathbb{R}^n_p)$ that is to a unitary operator related to $L$
as follows
\begin{equation}
\label{10}
\begin{array}{c}
U^{-1}\hat pU=L_1\hat P+L_2\hat Q,\\
U^{-1}\hat qU=L_3\hat P+L_4\hat Q.
\end{array}
\end{equation}
The operator $U$ can be represented as an integral operator
\begin{equation}
\label{11} (Uf)(p)=\int U(p,P)f(P)dP.
\end{equation}
with the kernel
\begin{equation}
\label{12} U(p,P)=C\exp\left[-\frac{i}{\hbar}S(p,P)\right],
\end{equation}
where the normalization constant
$C=((2\pi\hbar)^3|\det\,L_2|)^{-1/2}$.
Then using the unitary of the operator $U$ we can express the
initial wave function of the quantum Hamiltonian in terms of the
final wave functions by the following formula
\begin{equation}
\label{13}
\psi(p)=C\int\tilde\psi(P)\exp\left[-\frac{i}{\hbar}S(p,P)\right]dP,
\end{equation}
here $\psi(p)$ is the wave function of the original Hamiltonian and
$\tilde\psi(P)$ is the wave function of the unitary equivalent Hamiltonian.

\section{Diagonalization of quadratic Hamiltonians}
The purpose of this section is to demonstrate an algorithm  of
reducing the quadratic Hamiltonian
\begin{equation}
\label{14} H(p,q)=\frac{1}{2}\left(
\begin{array}{c}
p \\ q
\end{array}
\right)\tilde H \left(
\begin{array}{c}
p \\ q
\end{array}
\right)\,,
\end{equation}
where $\tilde H$ is a symmetric matrix of order $2n\times 2n$,
 to the canonical form. This
form must contain only squares of the corresponding momenta and
positions.

Then Hamilton equations corresponding to our quadratic Hamiltonian
(\ref{14}) can be written in the form $\dot x=I\tilde Hx$, here
$I$ is the symplectic unity
\begin{equation}
\label{15}
I=\left(
\begin{array}{cc}
0 & -E\\
E& 0
\end{array}
\right).
\end{equation}

At first we find the family $(\lambda_i)_{1\leq i\leq 2n}$ of
eigenvalues the matrix $I\tilde H$,
where $I$  is the symplectic unity
\begin{equation}
\label{15}
I=\left(
\begin{array}{cc}
0 & -E\\
E& 0
\end{array}
\right)\,,
\end{equation}
and the family $(f_i)_{1\leq i\leq 2n}$ of the corresponding
eigenvectors. For this purpose we use the following simple lemma

{\bf  Lemma.} {\it The characteristic polynomial $\det(I\tilde
H-\lambda E)$ has only real coefficients and contains only terms
of even degree}.

It follows from the Lemma that if $\lambda_i$ is an eigenvalue of
the matrix $IH$ then $-\lambda_i$ and $\lambda_i^*$ are
eigenvalues of this matrix too. Taking into account this assertion
we can arrange $\lambda_i$ in a finite sequence of the form
\begin{equation}
\label{17}
\nu_1,..,\nu_l,i\omega_1,..i\omega_m,-\nu_1,..,-\nu_l,-i\omega_1,..,-i\omega_m,
\end{equation}
where $\nu_i,\omega_k>0$; $l,m\geq0$; $l+m=n$.

Now we transform the basis $(f_i)$ into a symplectic one. Let
$(b_i)_{1\leq i\leq 2n}$ be the following sequence of vectors from
 ${\bf  R}^{2n}$:
\begin{equation}
\label{18} f_1(\nu_1),..,\ f_l(\nu_l), \ {\rm
Re}[f_1(i\omega_1)],..,\ {\rm Re}[f_m(i\omega_m)],\ f_1(-\nu_1),
..,\ f_l(-\nu_l),\ {\rm Im}[f_1(i\omega_1)],..,\ {\rm
Im}[f_m(i\omega_m)].
\end{equation}
It is clear that the vectors $(b_i)_{1\leq i\leq 2n}$ form a basis
of the space  ${\bf R}^{2n}$. These vectors satisfy the
skew-orthogonality conditions
\begin{equation}
\label{19} \begin{cases}[ b_i|b_k]=0, {\rm \ \ if\ \ } i<k;\ \
i+n\neq k,\cr [ b_i|b_{i+n}]\neq0, \ \ \forall \ i=1,..,n.
\end{cases}
\end{equation}
Denote $c_i=[ b_i|b_{i+n}]$,
then according to (\ref{19}) the vectors $b_i'$,
\begin{equation}
 \label{21}
\begin{cases}
\displaystyle b_i'=-\frac{{\rm sgn}(c_i)}{\sqrt{|c_i|}}\,b_i;\ \
i=1,..,n, \cr \displaystyle
b_{i+n}'=\frac{1}{\sqrt{|c_i|}}\,b_{i+n};\ \ i=1,..,n,
\end{cases}
\end{equation}
form a symplectic basis of the space ${\bf R}^{2n}$. Using
coordinates with respect to this basis, we reduce  the Hamiltonian
$H$ to the form
\begin{equation}
\label{22}
H=-2\sum_{i=1}^{l}\nu_iP_iQ_i+\sum_{i=1}^{m}\varepsilon_i\omega_i(P_{i+l}^2+
Q_{i+l}^2),
\end{equation}
where $\varepsilon_i={\rm sgn}\ c_i$.

Now we define the canonical basis of the quadratic form $\tilde H$
by the relations
\begin{equation}
\label{23} \begin{cases} a_i=b_i', {\rm \ \ if\ \ } l+1\leq i\leq
n {\rm \ \ and\ \ } n+l+1\leq i\leq 2n,\cr
a_i=-\frac{1}{\sqrt{2}}(b_i'+b_{i+n}'), {\rm\ \ if\ \ }1\leq i\leq
l,\cr a_{i+n}=\frac{1}{\sqrt{2}}(b_i'-b_{i+n}'), {\rm\ \ if\ \
}1\leq i\leq l.
\end{cases}
\end{equation}
With respect to this basis, the Hamiltonian $H$ has the canonical
form
\begin{equation}
\label{24}
H(P,Q)=\sum_{i=1}^{l}\nu_i\left(P_i^2-Q_i^2\right)+
\sum_{i=1}^{m}\varepsilon_i\omega_i\left(P_{i+l}^2+Q_{i+l}^2\right).
\end{equation}
The transition matrix from the initial phase coordinates $p,q$ to
the final ones $P,Q$ can be written in the following form
\begin{equation}
\label{25} L=[a_1,a_2,a_3,a_4,a_5,a_6]\,,
\end{equation}
where each vector $a_i$ is written in the corresponding column.
\section{Diagonalization of the Hamiltonian of a $3D$ anisotropic oscillator
with a magnetic field} Let us consider the Hamiltonian of a $3D$
anisotropic harmonic oscillator with a magnetic field $\bf B$
arbitrarily directed with respect to the potential symmetry axes
\begin{equation}
\label{26} H=\frac{1}{2m}\left({\bf  p}-\frac{e}{c}{\bf
A}\right)^2+\frac{m}{2}\left(
\Omega_x^2x^2+\Omega_y^2y^2+\Omega_z^2z^2\right).
\end{equation}
Here $\bf  A$ is the vector potential of the magnetic field, $c$
is the light velocity, $\Omega_i$ ($i=x,y,z$) are the
characteristic frequencies of the parabolic potential, $m$, $e$,
and $\bf p$ are the mass, the charge, and the momentum of the
considered particle.

The results of Sec.~III show that Hamiltonian (\ref{26}) is
unitary equivalent to a quadratic Hamiltonian without magnetic
field but with new frequencies named "hybrid frequencies". The
purpose of this section is to reduce Hamiltonian (\ref{26}) to the
canonical form
\begin{equation}
\label{27} H({\bf  P},{\bf
Q})=\frac{1}{2m}\left(P_1^2+P_2^2+P_3^2\right)+
\frac{m}{2}\left(\omega_1^2Q_1^2+\omega_2^2Q_2^2+\omega_3^2Q_3^2\right),
\end{equation}
using the algorithm developed in the preceding section; here ${\bf
P}, {\bf  Q}$ are the new phase coordinates, $\omega_i$
($i=1,2,3$) are the hybrid frequencies. It is convenient to choose
the following gauge for the  vector potential $\bf A$
\begin{equation}
\label{28} {\bf
A}=\left(\frac{1}{2}B_yz-B_zy,0,B_xy-\frac{1}{2}B_yx\right).
\end{equation}
Below we use the  notations
$$
K_1^2=m^2\Omega_x^2,\ \ K_2^2=m^2\Omega_y^2,\ \
K_3^2=m^2\Omega_z^2;
$$
$$
p_x=p_1,\ p_y=p_2,\ p_z=p_3;\ x=q_1,\ y=q_2,\ z=q_3;
$$
$$
\frac{eB_x}{2c}=B_1,\ \frac{eB_y}{2c}=B_2,\ \frac{eB_z}{2c}=B_3.
$$
In this notation
\begin{equation}
\label{32}
H=\frac{1}{2m}\left[(p_1-B_2q_3+2B_3q_2)^2+p_2^2+(p_3+B_2q_1-2B_1q_2)^2+
K_1^2q_1^2+K_2^2q_2^2+K_3^2q_3^2 \right],
\end{equation}
i.e., $H$ is a quadratic form of $p$ and $q$ with the matrix
\begin{equation}
\label{34}
\tilde H=\frac{1}{m}\left(\begin{array}{cccccc}
1&0&0&0&2B_3&-B_2\\
0&1&0&0&0&0\\
0&0&1&B_2&-2B_1&0\\
0&0&B_2&\bar K_1&-2B_1B_2&0\\
2B_3&0&-2B_1&-2B_1B_2&\bar K_2&-2B_2B_3\\
-B_2&0&0&0&-2B_2B_3&\bar K_3
\end{array}\right),
\end{equation}
where $\bar K_1=K_1^2+B_2^2$, $\bar K_2=K_2^2+4B_1^2+4B_3^2$,
$\bar K_3=K_3^2+B_2^2$.

To find the eigenvalues of the matrix $I\tilde H$ we solve the
equation $\det(I\tilde H-\lambda E)=0$, which is easily reduced to
the form
\begin{equation}
\label{35}
\left|\begin{array}{cccccc}
1&0&0&-\mu&2B_3&-B_2\\
0&1&0&0&-\mu&0\\
0&0&1&B_2&-2B_1&-\mu\\
0&0&0&K_1^2+\mu^2&-2B_3\mu&2B_2\mu\\
0&0&0&2B_3\mu&K_2^2+\mu^2&-2B_1\mu\\
0&0&0&-2B_2\mu&2B_1\mu&K_3^2+\mu^2
\end{array}\right|=0,
\end{equation}
where $\mu=m\lambda$. Since
\begin{equation}
\label{36}
\det\left(\begin{array}{cc}
A&B\\
0&C
\end{array}\right)=\det(AC)
\end{equation}
for any $n\times n$ matrix $A,B,C$,  we can obtain the hybrid
frequencies from the following sixth-order algebraic equation
[\onlinecite{Li91}]
\begin{equation}
\label{37}
(\Omega_x^2+\lambda^2)(\Omega_y^2+\lambda^2)(\Omega_z^2+\lambda^2)
+\omega_{xc}^2(\Omega_x^2+\lambda^2)\lambda^2
+\omega_{yc}^2(\Omega_y^2+\lambda^2)\lambda^2
+\omega_{zc}^2(\Omega_z^2+\lambda^2)\lambda^2=0,
\end{equation}
where $\omega_{ic}=eB_i/mc$ are the components of the cyclotron
frequency and $\lambda_i^{\pm}=\pm i\omega_i$. Note that equation
(\ref{37}) always has three different real negative solutions with
respect to $\lambda^2$.

Now our purpose is to find the eigenvectors of the matrix $I\tilde
H$ associated with the eigenvalues $\lambda_i^{\pm}$. Using
elementary matrix transformations one can reduce the matrix
$M=I\tilde H-\lambda E$ to the form
\begin{equation}
\label{38}
M=\frac{1}{m}\left(\begin{array}{cccccc}
0&0&0&-K_1^2-\mu^2&2B_3\mu&-B_2\mu\\
0&0&0&-2B_3\mu&-K_2^2-\mu^2&2B_1\mu\\
0&0&0&2B_2\mu&-2B_1\mu&-K_3^2-\mu^2\\
1&0&0&-\mu&2B_3&-B_2\\
0&1&0&0&-\mu&0\\
0&0&1&B_2&-2B_1&-\mu
\end{array}\right)\equiv\frac{1}{m}\left(
\begin{array}{cc}
0&X\\
E&Y
\end{array}
\right).
\end{equation}
The eigenvectors of (\ref{38}) are obtained from the system
\begin{equation}
\label{41} \begin{cases}X{\bf  q}=0,\cr {\bf  p}=-Y{\bf  q}.
\end{cases}
\end{equation}
Solving this system we get the coordinates of the eigenvectors
$g(\mu)$ associated with the eigenvalues $\mu=m\lambda^\pm_i$ of
$I\tilde H$:
\begin{equation}
\begin{array}{c}
p_1=-2B_3K_1^2(K_3^2+\mu^2)-4B_2^2B_3\mu^2-
2B_1B_2\mu(K_1^2-\mu^2),\\
p_2=\mu(K_1^2+\mu^2)(K_3^2+\mu^2)+4B_2^2\mu^3,\\
p_3=2B_1K_3^2(K_1^2+\mu^2)+4B_1B_2^2\mu^2-2B_2B_3\mu(K_3^2-\mu^2).
q_1=2B_3\mu(K_3^2+\mu^2)+4B_1B_2\mu^2,\\
q_2=(K_1^2+\mu^2)(K_3^2+\mu^2)+4B_2^2\mu^2,\\
q_3=4B_2B_3\mu^2-2\mu B_1(K_1^2+\mu^2)\,.\\
\end{array}
\end{equation}
Consider the vectors
$$
f_1={\rm Re\ }[g(-im\omega_1)],\ \ f_2={\rm Re\
}[g(-im\omega_2)],\ \ f_3={\rm Re\ }[g(-im\omega_3)],
$$
$$
f_4={\rm Im\ }[g(-im\omega_1)],\ \ f_5={\rm Im\
}[g(-im\omega_2)],\ \ f_6={\rm Im\ }[g(-im\omega_3)].
$$
Then we have for the coordinates of $f_i$ the following
expressions. If $i=1,2,3$, then
$$
 {\rm
Re}p_1=-2B_3K_1^2(K_3^2-m^2\omega_i^2)+4B_2^2B_3m^2\omega_i^2,\
{\rm Re}p_2=0,\
$$
$$
{\rm Re}p_3=2B_1K_3^2(K_1^2-m^2\omega_i^2)-4B_1B_2^2m^2\omega_i^2,\\
$$
$$
{\rm Re}q_1=-4B_1B_2m^2\omega_i^2,\ {\rm
Re}q_2=(K_1^2-m^2\omega_i^2)(K_3^2-m^2\omega_i^2)-4B_2^2m^2\omega_i^2,\
{\rm Re}q_3=-4B_2B_3m^2\omega_i^2.
$$
If $i=4,5,6$, then
$$
{\rm Im}p_1=2B_1B_2m\omega_{i-3}(K_1^2+m^2\omega_{i-3}^2),\ {\rm
Im}p_2=-m\omega_{i-3}(K_1^2-m^2\omega_{i-3}^2)(K_3^2-m^2\omega_{i-3}^2)
+4B_2^2m^3\omega_{i-3}^3,\
$$
$$
{\rm Im}p_3=-2B_2B_3m\omega_{i-3}(K_3^2+m^2\omega_{i-3}^2),
$$
$$
{\rm Im}q_1=-2B_3m\omega_{i-3}(K_3^2-m^2\omega_{i-3}^2),\ {\rm
Im}q_2=0,\ {\rm Im}q_3=2m\omega_{i-3}
B_1(K_1^2-m^2\omega_{i-3}^2).
$$


Denote
$$
M_i=m^4\left\{
\omega_{yc}^2\omega_i^4(\omega_{xc}^2+\omega_{yc}^2)+
\omega_{xc}^2\Omega_z^2(\Omega_x^2-\omega_i^2)^2+\omega_{zc}^2\Omega_x^2
(\Omega_z^2-\omega_i^2)^2+ \right.
$$
\begin{equation}
+\left. \left[
(\Omega_x^2-\omega_i^2)(\Omega_z^2-\omega_i^2)-\omega_{yc}^2\omega_i^2
\right]^2 \right\}^{1/2}
\end{equation}
for $i=1,2,3$; and $M_i=M_{i-3}$ for $i=4,5,6$. It is easy to show
that the following basis is  symplectic
\begin{equation}
h_i=\frac{1}{M_i\sqrt{m\omega_i}}f_i\,,
\end{equation}
where we put $\omega_i=\omega_{i-3}$ for $i=4,5,6$.

With respect to this basis the Hamiltonian $H$ has the form
\begin{equation}
 H=\frac{1}{2}\sum_{i=1}^{3}\omega_i(\tilde p_i^2+\tilde q_i^2),
\end{equation}
here $\tilde p_i$, and $\tilde q_i$ are new phase coordinates.

To reduce this Hamiltonian to the canonical form (\ref{27}) we
change the variables once again
\begin{equation}
P_i=\sqrt{m\omega_i}\tilde p_i;\ \
Q_i=\frac{1}{\sqrt{m\omega_i}}\tilde q_i.
\end{equation}

Finally, the transition matrix $L$ from the initial coordinates to
the coordinates $P_i$, $Q_i$ has the  components
\begin{equation}
l_{ij}=\frac{a_{ij}}{M_i\sqrt{m\omega_i}},
\end{equation}
where $a_{ij}$ are defined as follows:

if $i=1,2,3$ then
$$
a_{i1}=-m^5\omega_{zc}\Omega_x^2(\Omega_z^2-\omega_i^2)+
\frac{1}{2}m^5\omega_{zc}\omega_{yc}^2\omega_i^2,\ \ a_{i2}=0,\ \
a_{i3}=m^5\omega_{xc}\Omega_z^2(\Omega_x^2-\omega_i^2)-
\frac{1}{2}m^5\omega_{xc}\omega_{yc}^2\omega_i^2,
$$
$$
a_{i4}=-m^4\omega_{xc}\omega_{yc}\omega_i^2,\ \
a_{i5}=m^4(\Omega_x^2-\omega_i^2)(\Omega_z^2-\omega_i^2)-
m^4\omega_{yc}^2\omega_i^2,\ \
a_{i6}=-m^4\omega_{yc}\omega_{zc}\omega_i^2;
$$
if $i=4,5,6$ then
$$
a_{i1}=\frac{1}{2}m^5\omega_{xc}\omega_{yc}\omega_{i-3}(\Omega_x^2+\omega_{i-3}^2),\
\
a_{i2}=-m^5\omega_{i-3}(\Omega_x^2-\omega_{i-3}^2)(\Omega_z^2-\omega_{i-3}^2)+
m^5\omega_{yc}^2\omega_{i-3}^3,
$$
$$
a_{i3}=\frac{1}{2}m^5\omega_{yc}\omega_{zc}\omega_{i-3}(\Omega_z^2+\omega_{i-3}^2),\
\ a_{i4}=-m^4\omega_{zc}\omega_{i-3}(\Omega_z^2-\omega_{i-3}^2),
$$
$$
a_{i5}=0,\ \
a_{i6}=m^4\omega_{xc}\omega_{i-3}(\Omega_x^2-\omega_{i-3}^2).
$$
As a result we have obtained an explicit formula for the
transition matrix $L$.

Note that the quantum counterpart of classical Hamiltonian
(\ref{22}) is the quantum harmonic oscillator with  the
eigenvalues
\begin{equation}
                     \label{eig}
E_{n_1n_2n_3}=\hbar\omega_1\left(n_1+\frac{1}{2}\right)
+\hbar\omega_2\left(n_1+\frac{1}{2}\right)
+\hbar\omega_3\left(n_1+\frac{1}{2}\right),
\end{equation}
where $n_1,n_2,n_3=0,1,..$.  In the $P$-representation, the
corresponding eigenfunctions are
\begin{equation}
\label{b18} \tilde\psi_n({\bf
P})=(\omega_1\omega_2\omega_3)^{-1/4}
\chi_{n_1}\left(\frac{P_1}{\sqrt{\omega_1\hbar}}\right)
\chi_{n_2}\left(\frac{P_1}{\sqrt{\omega_1\hbar}}\right)
\chi_{n_3}\left(\frac{P_1}{\sqrt{\omega_1\hbar}}\right)\,.
\end{equation}
Here $\chi_n$ is the $n$-th oscillator function
$$
\chi_n(x)=(\pi^{1/2}2^nn!)^{-1/2}\exp(-x^2/2)H_n(x)\,,
$$
where $H_n$ is the corresponding Hermite polynomial.

\section{Calculating the Berry phase}
The purpose of this section is to obtain an explicit formula for
the Berry phase of a $3D$ anisotropic parabolic quantum well
placed in a precessing magnetic field. We deal with the case when
a considered quantum state undergoes an adiabatic evolution in the
space of the values of the magnetic induction $\bf B$ in such a
way that  eigenvalues (\ref{eig}) remain nondegenerate. Let
$\psi_{\bf n}$ be the wave function of the original state (here
${\bf n}=(n_1,n_2,n_3$) is the set of the quantum numbers). If the
system undergoes an evolution in question, then the wave function
obtains the geometric phase
\begin{equation}
\label{b1} \gamma_{\bf n}=\oint\limits_{C}{\bf V_n}({\bf B})d{\bf
B},
\end{equation}
where $C$ is a closed path ${\bf  B}(t)$ in the parameter space
such that ${\bf  B}(T)={\bf  B}(0)$, and ${\bf  V_n}({\bf B})$ is
so-called Berry vector potential given by
\begin{equation}
\label{b2} {\bf V_n}({\bf B})=i\langle\psi_n({\bf B})|\nabla_{{\bf
B}}\psi_n({\bf B})\rangle=-{\rm Im }\langle\psi_n({\bf
B})|\nabla_{{\bf B}}\psi_n({\bf B})\rangle.
\end{equation}
Since the Berry phase is determined by ${\bf V_n}({\bf B})$,  it
is sufficient to calculate  only the vector potential. In our case
the magnetic field doesn't vanish and hence the parameter space is
 ${\rm\bf R}^3\backslash \{0\}$, which is topologically
non-trivial, and we can expect that the Berry phase is
non-trivial, too.

To find the scalar product in (\ref{b2}) one needs eigenfunctions
of the initial Hamiltonian. However, a direct calculation of the
wave functions to obtain the Berry phase is a complicated
computational problem. To calculate the Berry phase, we suggest to
simplify the calculations using the method of linear canonical
transformation of the phase space developed in the preceding
sections. In this case we can use formula (\ref{11}) to express
the initial wave functions
$\psi_{\mathbf{n}}(\mathbf{P};\mathbf{B})$ in terms of the final
ones (\ref{b18}):

\begin{equation}
 \label{b3}
 \psi_{\bf n}({\bf  p};{\bf  B})= C\int{\tilde
\psi_{\bf n}}({\bf P};{\bf  B})\exp\left[ -\frac{i}{\hbar}S({\bf
p},{\bf P})\right]d{\bf  P}\,.
\end{equation}

Taking into account that ${\tilde \psi_{\bf n}}({\bf P};{\bf B})$
is real-valued we can calculate $\nabla_{\bf B}{\psi_{\bf n}}({\bf
p};{\bf B})$:
$$
\nabla_{{\bf B}}\psi_{\bf n}({\bf p};{\bf B})=\int\left[
\nabla_{{\bf B}} C\tilde{\psi}_n({\bf P};{\bf B})\right]\exp\left[
-\frac{i}{\hbar}S({\bf p},{\bf B}) \right]d\mathbf{P}-
$$
\begin{equation}
\label{b4} -\frac{i}{\hbar}\int C\tilde{\psi}_{\bf n}({\bf P;{\bf
B}})\exp\left[ -\frac{i}{\hbar}S({\bf p},{\bf P}) \right]\left[
\nabla_{{\bf B}} S({\bf p},{\bf P}) \right]d{\bf P}.
\end{equation}
In what follows we denote $\nabla_{\bf B}$ simply by $\nabla$.

Let us calculate scalar product  (\ref {b2})
$$
\langle\psi_{\bf n}|\nabla\psi_{\bf n})\rangle= \int d{\bf
p}\psi_{\bf n}^*({\bf p})\nabla\psi_{\bf n}({\bf p})= \int d{\bf
p}\left\{ C^*\int\int d{\bf P}d{\bf P}'\tilde{\psi}_{\bf n}({\bf
P}') \left[ \nabla C\tilde{\psi}_{\bf n}({\bf P}) \right]\times
\right.
$$
$$
\times\exp\left[ \frac{i}{\hbar}\left(S({\bf p},{\bf P}')-S({\bf
p},{\bf P})\right) \right]-\frac{i}{\hbar} \int\int d{\bf P}d{\bf
P}'|C|^2\tilde{\psi}_{\bf n}({\bf P}')\tilde{\psi}_{\bf n}({\bf
P})\times
$$
\begin{equation}
\label{b5}\times\left.\exp\left[ \frac{i}{\hbar}\left(S({\bf
p},{\bf P}')-S({\bf p},{\bf P})\right) \right]\nabla S({\bf
p},{\bf P})\right\}.
\end{equation}
By changing the variables $L_2^{-1}{\bf p}/\hbar={\bf y};\ \ d{\bf
p}=|\det{L_2}|\hbar^3d{\bf y}$, we perform the integration with
respect to $\mathbf{p}$ in the first term of (\ref{b5}) and get
$$
\int\exp\left[
\frac{i}{\hbar}(S({\bf p},{\bf P}')-S({\bf p},{\bf P}))
\right]d{\bf p}=
$$
$$
=\exp\left[ \frac{i}{2\hbar}(\langle{\bf P}'|L_2^{-1}L_1{\bf
P}'\rangle-\langle{\bf P}|L_2^{-1}L_1{\bf P}\rangle)\right] |\det
L_2|\hbar^3\int\exp\left[i\langle{\bf P}-{\bf P}'|{\bf
y}\rangle\right]d{\bf y}=
$$
\begin{equation}
\label{b6}=(2\pi\hbar)^3|\det{L_2}|\delta({\bf P}-{\bf P}').
\end{equation}
Hence the first term in (\ref{b5}) is real-valued and doesn't
contribute in the Berry phase. Therefore, we need to calculate
only the second term in (\ref{b5}).

It is convenient to introduce the notations
\begin{equation}
\label{b7} A=-\nabla L_2^{-1};\ B=\frac{1}{2}\nabla(L_2^{-1}L_1);\
D=\frac{1}{2}\nabla(L_4L_2^{-1});\
\Lambda=\frac{1}{2\hbar}(\langle{\bf P}'|L_2^{-1}L_1{\bf
P}'\rangle-\langle{\bf P}|L_2^{-1}L_1{\bf  P}\rangle).
\end{equation}
Now integrating the second term in (\ref{b5}) with respect to $\bf
p$, we obtain
$$
I=\int\exp\left[\frac{i}{\hbar}(S({\bf p},{\bf P}')-S({\bf p},{\bf
P})) \right]\nabla S({\bf p},{\bf P})d{\bf p}=
$$
$$
\hbar^3|\det{L_2}|\exp\left(i\Lambda\right)\times
$$
$$
\left[ (2\pi)^3\langle{\bf P}|B{\bf P}\delta({\bf P}-{\bf
P}')\rangle-i\hbar(2\pi)^3\sum_{k=1}^{3} [(AL_2)^*{\bf
P}]_k\partial_k\delta({\bf P}-{\bf P}')-\right.
$$
\begin{equation}
\label{b10}
\left.\hbar^2(2\pi)^3(\langle{\bf\partial}|
L_2^*DL_2{\bf\partial}\rangle)\delta({\bf P}-{\bf P}') \right].
\end{equation}

Calculating the corresponding derivatives
\begin{equation}
\label{b12}
\partial_k\exp\left(i\Lambda
\right)= -\frac{i}{\hbar}\left[(L_2^{-1}L_1{\bf
P})_k\right]\exp\left(i\Lambda \right),
\end{equation}
\begin{equation}
\label{b13}
\partial_l\partial_k\exp\left(i\Lambda
\right)=-\frac{i}{\hbar}
(L_2^{-1}L_1)_{kl}\exp\left(i\Lambda\right)-\frac{i}{\hbar}
(L_2^{-1}L_1{\bf P})_k(L_2^{-1}L_1{\bf P})_l\exp\left( i\Lambda
\right),
\end{equation}
and substituting (\ref{b12}) and (\ref{b13}) into (\ref{b10})
after simple algebra we get, taking away the imaginary part,
$$
 \mathrm{Re} I=
$$
\begin{equation}
\label{b14}
(2\pi\hbar)^3|\det L_2|\left\{ \langle{\bf P}'|B{\bf
P}'\rangle+\langle{\bf P}'|AL_1{\bf P}'\rangle+\langle{\bf
P}'|L_1^*DL_1{\bf P}'\rangle -\hbar^2\langle{\bf
\partial}|L_2^*DL_2{\bf
\partial}\rangle\right\}\delta({\bf P}-{\bf P}').
\end{equation}
As a result we have the following formula for the Berry vector potential
$$
{\bf V_n}({\bf B})=-\mathrm{Im}\int d\mathbf{p}\psi_{\bf n}^*({\bf
p})\nabla\psi_{\bf n}({\bf p})= \frac{1}{\hbar}\int\int d{\bf
P}d{\bf P}|C|^2\tilde\psi_n({\bf P}')\psi_n({\bf P})(\mathrm{Re}
I)=
$$
$$
\frac{1}{\hbar}(2\pi\hbar)^3|\det L_2||C|^2\left\{\int d{\bf
P}'\left[ \langle{\bf P}'|B{\bf P}'\rangle+\langle{\bf
P}'|(AL_1){\bf P}'\rangle+\langle{\bf P}'|(L_1^*DL_1){\bf
P}\rangle \right]|\tilde\psi_n({\bf P}')|^2+\right.
$$
\begin{equation}
\label{b15} \left.\hbar^2\int d{\bf P}d{\bf
P}'\sum_{k,l=1}^3(L_2^*DL_2)\partial_l\partial_k\delta({\bf P}
-{\bf P}') \tilde\psi_n({\bf P}')\tilde\psi_n({\bf P})\right\}.
\end{equation}
Denote
\begin{equation}
\label{b16} \begin{cases} F=B+AL_1+L_1^*DL_1, \cr G=L_2^*DL_2.
\end{cases}
\end{equation}
It is clear that $G^*=G$ and we can rewrite (\ref{b15}) in the
form
$$
{\bf V_n}({\bf B})=\frac{1}{\hbar}(2\pi\hbar)^3|\det
L_2||C|^2\left\{\int d{\bf P}'\langle{\bf P}'|F{\bf
P}'\rangle|\tilde\psi_{\bf n}({\bf P}')|^2d{\bf P}'+\right.
$$
\begin{equation}
\label{b17} \left.\hbar^2\int d{\bf P}d{\bf
P}'\sum_{k,l=1}^3G_{lk}\partial_l\partial_k \delta({\bf P}-{\bf
P}') \tilde\psi_{\bf n}({\bf P}')\tilde\psi_{\bf n}({\bf
P})\right\}.
\end{equation}
Using (\ref{b18}), we get
$$
\int P_kP_l|\tilde\psi_n({\bf P})|^2d{\bf P}=0\,,
$$
if $k\ne l$, and
$$
\int P_k^2|\tilde\psi_n({\bf P})|^2d{\bf P}=
\omega_k\hbar\int\chi_{n_k}^2(q)q^2dq=\left(n_k+\frac{1}{2}\right)\hbar\omega_k\,.
$$

Since
\begin{equation}
\label{b20} \int\chi_n(x)\chi_n'(x)dx=0,
\end{equation}
it follows for $k\neq l$
\begin{equation}
\label{b21} \int\tilde\psi_n({\bf
P})\partial_k\partial_l\tilde\psi_n({\bf P})d{\bf P}=0\,.
\end{equation}
If $k=l$, then
\begin{equation}
\label{b22} \int\tilde\psi_n({\bf P})\partial_k^2\tilde\psi_n({\bf
P})d\mathbf{P}
=-\frac{1}{\omega_k\hbar}\left(n_k+\frac{1}{2}\right).
\end{equation}
Finally we have
\begin{equation}
\label{b23} {\bf V_n}({\bf B})=\sum_{k=1}^3\left( n_k+\frac{1}{2}
\right) \left[ \omega_kF_{kk}+\frac{G_{kk}}{\omega_k} \right].
\end{equation}
Equation (\ref{b23}) is the main result of our paper. This formula
extends a result of Berry [\onlinecite{Ber85}] to the $3D$ case
and contains as a particular case some results from
[\onlinecite{Fer92}].

\noindent{\bf Remark.} Note that Formula (\ref{b23}) is valid in
more general situations. Namely, let $H(\xi)$ be a family of
quadratic Hamiltonians in the state space $L^2(\mathbb{R}^3)$
depending on a parameter $\xi$, $\xi\in X$, where the parameter
space $X$ is a smooth manifold. Let $\xi=\xi(t)$ be a curve in $X$
such that $H(\xi(t))$ has three different eigenvalues at each $t$.
Then the Berry phase corresponding to an adiabatic evolution along
the curve $\xi=\xi(t)$ is given by (\ref{b23}), if we replace
$\nabla_\mathbf{{B}}$ by $\nabla_\xi$ in all the auxiliary
expressions.

\begin{acknowledgments}
 The authors are very grateful to J.Br\"uning and K.Pankrashkin for useful
 discussions.
 We gratefully acknowledge grants of DFG,
INTAS and RFBR (Projects No. 01-02-16564 and 02-01-00804). We are
also thankful to Humboldt University of Berlin for warm
hospitality during the preparation of this paper.
\end{acknowledgments}

\end{document}